\documentclass{article}
\begin{document}
\newcommand{\bmu}{\mbox{\boldmath$\mu$}}
\title{What is the force on a magnetic dipole?}
\author{Jerrold Franklin\footnote{Internet address:
Jerry.F@TEMPLE.EDU}\\
Department of Physics\\
Temple University, Philadelphia, PA 19122-1801}
\maketitle
\begin{abstract}
This paper will be of interest to physics graduate students and faculty.
We show that attempts to modify the force on a magnetic dipole by introducing either hidden momentum
 or internal forces are not correct.
The standard textbook result
${\bf F=\nabla(\bmu\cdot B)}$  is correct 
even in the presence of time dependent electromagnetic fields.
\end{abstract}

\section{Introduction}
There has been some controversy in the literature about what is the correct expression for the force on a magnetic dipole. 
Textbooks generally give the expression\footnote{We use Gaussian units with c=1.}
\begin{equation}
{\bf F=\nabla(\bmu\cdot B)}
\label{f1}
\end{equation}
for the force $\bf F$ on a magnetic dipole {\bmu} in a magnetic field $\bf B$.
This is derived, for example, for the magnetic dipole moment of a localized current density in \cite{j}\footnote{The third edition (1999) of this textbook repeats the derivation of the second edition
giving Eq.~(1) above, but then adds the statement
``The effective force in Newton's equation of motion of mass times acceleration is (5.69), augmented by $(1/c^2)(d/dt){\bf(E\times\bmu)}$.",
presumably basing this on the claims in \cite{v} and \cite{h}.}
 and \cite{b}, and for a small current loop in \cite{jf}.
On the other hand, two papers\cite{v,h} have given arguments that the force on a magnetic dipole should be
\begin{equation}
{\bf F=\nabla(\bmu\cdot B)}-\frac{d}{dt}(\bmu\times{\bf E}).
\label{f2}
\end{equation}

Which of these equations is the correct one? In this paper, we show that the arguments made in \cite{v} and \cite{h} are not correct, and the usual textbook equation (\ref{f1}) gives the correct force on a magnetic dipole.

\section{Force on a magnetic dipole}

The addition of the second term in Eq.~(\ref {f2}) was attributed to ``hidden momentum" in \cite{v},
with the additional force arising by subtracting the time derivative of the hidden momentum from the force given by Eq.~(\ref{f1}).
Hidden momentum was first suggested by Shockley and James\cite{sj} in 1967. A review of many of the subsequent papers using hidden momentum 
has been given by Griffiths\cite{g}.\footnote{We, however, have published an article\cite{jfajp} disputing the general occurrence of hidden momentum as a counterbalance to electromagnetic field momentum.}

In \cite{v}, Vaidman considerered three models of a magnetic dipole:\\
(i) A gas of charged particles constrained to move inside a {\it neutral} tube;\\
(ii) A gas of charged particles constrained to move inside a {\it conducting} tube;\\
(iii) A charged (incompressible) fluid constrained to move inside a {\it neutral} tube.\\
For some reason he did not consider a current in a conducting wire, which is the common form of a magnetic dipole.

The presumed occurrence of hidden momentum usually depends on an external electric field affecting the current.
For this reason, Vaidman considered that his case (ii) would have no hidden momentum because the electric field would not penetrate the conducting tube. However, the same is true of model (i) because it has free charges inside the tube. If an electric field penetrated the tube, the free charges would move to form a surface charge that would reduce the electric field inside the tube to zero. This is just the same mechanism that produces zero electric field inside a conducting tube, or a conducting wire. Thus there is no essential difference between his models (i), (ii), and a conducting wire. They all  have no hidden momentum, 
so hidden momentum cannot be relied upon to produce the second term in Eq.~(\ref{f2}) for those cases.

We are left with Vaidman's case (iii), a charged (incompressible) fluid constrained to move inside a neutral tube.
This case cannot have standard hidden momentum because every charge in the incompressible fluid must move at the same velocity.
Instead, Vaidman attributes hidden momentum in this case to mechanical stress within the incompressible fluid.

He first relates the pressure in a stationary fluid to the electric potential $\phi$ of an external electric field as
\begin{equation}
p=-\rho\phi,
\label{pr}
\end{equation}
where $\rho$ is the charge density in the fluid.
He asserts that this leads to an energy momentum tensor with pressure terms given by
\begin{equation}
T^{11}=T^{22}=T^{33}=-\rho\phi.
\label{T3}
\end{equation}

He then states that ``Lorentz transformation of the energy momentum tensor shows that these pressure terms yield in the rest frame of the tube, the off-diagonal momentum density terms:
\begin{equation}
T^{i0}=-(v_i/c)\rho\phi=(-1/c)J_i\phi."
\label{T3i0}
\end{equation}
He goes on to show that this momentum density leads to the hidden momentum he needs to deduce the second term of Eq.~(\ref{f2}).

The problem with this derivation is that Vaidman has left out the full structure of the four dimensional energy momentum tensor for a fluid at rest. In addition to the three terms he gave in Eq.~(\ref{T3}) above, there is a term
\begin{equation}
T^{00}=\rho\phi,
\label{T00}
\end{equation}
corresponding to the energy density in the fluid at rest.
Thus the full structure of the energy momentum tensor for the fluid at rest is
\begin{equation}
[{\bf T}]=\rho\phi[\bf G],
\label{T4}
\end{equation}
where $[{\bf G}]$ is the four dimensional Minkowski metric tensor, which is diagonal and idempotent, and thus is diagonal in any Lorentz frame.
Therefore, Lorentz transforming the correct four dimensional energy momentum tensor does not lead to off-diagonal
momentum density terms $T^{i0}$ in any frame.

We have now shown that there is no hidden momentum
in any of the three models proposed by Vaidman, so his basic premise is wrong.
Since Vaidman realized that his model (ii) does not have hidden mometum, he had to conjecture the addition of the second term in Eq. (\ref{f2}), and challenged ``the reader
to provide a proof for an arbitrary shape of the conductor in model (ii)". 
 
Vaidman's challenge to the reader was taken up by Hnizdo in \cite{h}, but his derivation of the second term in Eq.~(\ref{f2}) was fatally flawed. He attempted to derive the second term for Vaidman's model (ii)  by introducing internal forces that would act on the magnetic dipole. 
He suggested that a changing external electric field would induce currents, ${\bf j}_{\rm induced}$, and that these induced currents would lead to two forces:
A force $\bf F_j$ caused by  the magnetic field of the induced currents
acting on the original current $\bf j$ of the magnetic moment,
and a force ${\bf F}_q$ due to the magnetic field $\bf B_j$ of the original current acting on the new induced currents.
These two forces would be given by
\begin{eqnarray}
{\bf F_j}&=&\int{\bf j\times B}_{\rm induced}d^3r,
\label{fj}\\
{\bf F}_q&=&\int{\bf j}_{\rm induced}\times{\bf B_j}d^3r.
\label{fq}
\end{eqnarray}

He then showed, after some algebra, that the sum of these two forces could be written as [Eq.~(5) of \cite{h}]
\begin{equation}
{\bf F}_{\rm internal}=
{\bf F_j+F}_q=
\int d^3r{\bf j(r})\frac{d}{dt}\phi_{\rm induced}({\bf r}),
\label{fin}
\end{equation}
where $\phi_{\rm induced}$ is the electrostatic potential due to the induced surface charges on the dipole.

Hnizdo took the force ${\bf F}_{\rm internal}$
to be a force acting on the dipole that would move the dipole if not balanced by an external force.
However, it can be seen from Eq.~(\ref{fin}) that the differential force on the induced charges is everywhere parallel to the original current in the dipole,
and therefore parallel to the surface of the conductor carrying the current.
Because of this, the force on the charges will act to change the surface charge distribution on the dipole, but will not actually be a push on the wire carrying the current.
In fact, it is this force that rearranges the surface charges so that they will continue to keep the electric field out of the interior.

Only the component of a force perpendicular to the surface of a conductor would cause a motion of the conductor.  This is because the surface charges cannot move away from the surface, since this would allow the external electric field to enter the conductor. If the direction of the force is parallel to the surface of the conductor, as in this case, it will cause charges to move parallel to the surface, but not produce a force on the conductor.

A third internal force, $\bf F_{\bmu}$, proposed by Hnizdo would be
due to an electric field induced by the magnetic field of a time varying magnetic dipole. The suggestion was that this field would produce a force on the surface charges that had originally been induced by the external electric field.
But this force too would be parallel to the surface of the conductor and only move the surface charges along the surface
without producing a force on the dipole.

 We have now shown that none of the mechanisms proposed in \cite{v} or \cite{h}
lead to the addition of the term $-\frac{d}{dt}(\bmu\times{\bf E})$
 for any of the models proposed by Vaidman.
Neither Vaidman nor Hnizdo considered the cases of intrinsic magnetic dipoles or ferromagnetic dipoles. Neither of these cases would have the hidden momentum proposed by Vaidman or the internal currents used by Hnizdo, so they too would satisfy the force law of
Eq.~(1) without the added term in Eq.~(2).

One other argument given by Vaidman in \cite{v} relates to the paper by Shockley and James\cite{sj}, which considered point charges in the presence of a time varying magnetic moment  (two counter-rotating oppositely charged discs in their example). They claimed that without a force on the rotating discs
there would be a force on the point charges, but no balancing force (equal and opposite) on the discs to preserve conservation of momentum.
Their example was treated in more detail by Coleman and Van Vleck\cite{cv} using a semi-relativistic Darwin Lagrangian
for a system consisting of the rotating charged discs and a point charge, along with the electromagnetic fields of the discs and the point charge.

In each case, a rate of change of momentum was found in addition to the rate of change of momentum of the point charge.
This was interpreted in each paper as a mechanical force on the charged discs corresponding to a change in the proposed hidden momentum in the discs.  However, the correct interpretation is that the momentum change was the rate of change of the electromagnetic momentum that is known to exist in the presence of electric and magnetic fields.  

An easy way to see that there will be no mechanical force on the oppositely charged discs is that the simple Lorentz force on the charges on the discs is given by $\frac{\bf dp}{dt}=q{\bf E}$, where $\bf E$ is the electric field due to the point charge.  This force just cancels because of the opposite charges on the discs, so there can be no mechanical force on the system of the two discs.  

When electromagnetic momentum is considered along with mechanical momentum, the sum of the mechanical momentum and electromagnetic momentum is constant.
In fact, the lack of conservation of purely mechanical momentum is what is responsible for the production of the electromagnetic momentum. This point is emphasized in the introduction to \cite{jfajp}, and it is also how the concept of electromagnetic momentum is generally introduced in textbooks.

It is the change in this electromagnetic momentum that balances the force on the point charge to preserve conservation of momentum.
Shockley and James calculated the rate of change of the momentum due to the combined electromagnetic fields of a point charge and rotating charged discs, but this does not correspond to a force on the discs.

\section {Conclusions}

Our conclusions are:
\begin{enumerate}
\item
 All of the attempts in \cite{v} and \cite{h} to produce the second term $-\frac{d}{dt}(\bmu\times{\bf E})$ in Eq.~(\ref{f2}) are wrong. 
The standard result ${\bf F=\nabla(\bmu\cdot B)}$ that has appeared in many textbooks for many years is still correct, 
so this article will have a positive impact on graduate education.
\item There is no hidden momentum in any of the magnetic dipole - electric field configurations of [4].

\end{enumerate}

 \end{document}